\documentclass[aps,twocolumn,superscriptaddress,floatfix,prl,10pt]{revtex4-2}

\usepackage{graphicx}
\usepackage[utf8]{inputenc}
\usepackage{amssymb}
\usepackage{amsmath}
\usepackage[colorlinks,allcolors=blue]{hyperref}
\usepackage{braket}
\usepackage{placeins}

\begin{document}

\setlength\abovedisplayskip{5pt}
\setlength\belowdisplayskip{5pt}
\title{Spectrum statistics in the integrable Lieb-Liniger model}
\author{Samy Mailoud}
\affiliation{Instituto de F\'{i}sica, Benem\'{e}rita Universidad Aut\'{o}noma
  de Puebla, Apartado Postal J-48, Puebla 72570, Mexico}
\author{Fausto Borgonovi}
\affiliation{Dipartimento di Matematica e
  Fisica and Interdisciplinary Laboratories for Advanced Materials Physics,
  Universit\`a Cattolica, via Musei 41, 25121 Brescia, Italy}
\affiliation{Istituto Nazionale di Fisica Nucleare,  Sezione di Pavia,
  via Bassi 6, I-27100,  Pavia, Italy}
\author{Felix M. Izrailev}
\affiliation{Instituto de F\'{i}sica, Benem\'{e}rita Universidad Aut\'{o}noma
  de Puebla, Apartado Postal J-48, Puebla 72570, Mexico}
\affiliation{Dept. of Physics and Astronomy, Michigan State University, E. Lansing, Michigan 48824-1321, USA}
 
\date{\today}

\begin{abstract}
We address the old and widely debated question of the statistical properties of integrable quantum systems, through the analysis of the paradigmatic Lieb-Liniger model. This quantum many-body model of 1--d interacting bosons allows for the rigorous determination of energy spectra via the Bethe ansatz approach and our interest is  understanding whether  Poisson statistics is a characteristic feature of this model. Using  both analytical and numerical studies we show that the properties of spectra strongly depend on whether the analysis is done for a full energy spectrum or for a single subset with fixed total momentum. We show that the Poisson distribution of spacing between nearest-neighbor energies can occur only for a set of energy levels with fixed total momentum, for neither too large nor too weak interaction strength, and for sufficiently high energy. On the other hand, when studying long-range correlations between energy levels, we found strong deviations from the predictions given  by a Poisson process. 


\end{abstract}

\maketitle

\section{Introduction} 
\label{s-I}

Since the first studies of one-body quantum systems that are strongly chaotic in the classical limit \cite{berry,valz,BGS,CCIF79,bill}, the most popular test widely used to distinguish between regular and chaotic systems was the search of the form of the nearest-neighbor level spacing distribution (LSD) for energy levels. Specifically, it was assumed that for a completely integrable system the LSD has generically the form of a Poisson distribution, $P(s)=\exp(-s)$, characterized by the absence of repulsion between close energies. Contrarily, in the opposite limit  of strong chaos, the LSD reflects strong repulsion of close energies, $P(s) \sim s^{\beta}$ (for $s \rightarrow 0$) with  $\beta = 1, 2, 4$ depending on the global symmetric properties of the system (see for example, \cite{R92}). Although it was shown that such a correspondence is not always exact \cite{WVFS90,BJS03,BJL03}, the counterexamples are often considered as quite specific ones.   

Historically, the interest in the properties of the LSD has been motivated by the experimental studies of quantum spectra of heavy nuclei and multi-electron atoms (for references, see, for example, \cite{FM09}). One of the first applied studies concerning the form of the LSD is traced back to 1939 \cite{G39}. Specifically, in view of the problem of phase transitions in  nuclear matter, it was assumed that the LSD has the form of a Poisson distribution. This and other experimental studies of the low energy neutron scattering in nuclear reactions has triggered intensive discussions of the typical form of the LSD (see in \cite{FM09}). After long discussions, it was accepted that according to some scaling arguments presented by Wigner \cite{W51}, the distribution $P(s)$ in application to heavy nuclei might be described by the expression nowadays known as the Wigner surmise (WS). The distinguishable difference of the WS from the Poisson distribution is the repulsion of nearest-neighbor energy levels, namely, $P(s) \rightarrow 0$ for $s \rightarrow 0$. As for the far tails of $P(s)$ for $s \rightarrow \infty$, the WS suggests an even  stronger decrease, $P(s) \sim \exp(-B s^2)$ if compared with the exponential decay $\sim \exp(-s)$ of the Poisson distribution.  This behavior has been confirmed in \cite{HH58} for the data gathered from neutron spectroscopy groups around the world, to obtain the first global spacing distribution of s-wave neutron resonances.  

Later on, following Wigner's studies of random matrices \cite{Wigner}, Dyson rigorously derived exact expressions for the tails of $P(s)$ for all values of $\beta$ \cite{Dyson}. According to these results, for $\beta = 1,4$, the tails are described by both exponential and Gaussian terms, and only for $\beta = 2$ the exponential term is absent. However, as was noted by Dyson himself, in the applications one can correctly resolve the tails of $P(s)$ only when the number of energy levels is very large, i.e. exceeding $10^{5}$. Clearly, this is not possible experimentally, therefore, the Wigner-Dyson (WD) expression $P(s) = A s^{\beta} \exp(-Bs^{2})$ (with $A, B$ being normalization constants) can be used as a good approximation in many applications.

Numerous experimental data obtained in the study of energy spectra of heavy nuclei, complex atoms, and molecules have confirmed the  emergence of the WD distribution (see, for example, references in \cite{GMW98}). It was, however, understood that the absence of level repulsion does not necessarily mean  absence of strong chaos. The point is that in the  analysis of  experimental or numerical data one has to be sure that the considered energy spectrum does not contain any  subset associated with some specific quantum numbers. Indeed, since such subsets are independent one from each other, the energy levels associated with different quantum numbers turn out to be completely  uncorrelated thus giving rise to the apparent absence of level repulsion. Thus, by superimposing the subsets of energies belonging to different quantum numbers, one can get a LSD which may not show any repulsion at all, while when fixing all quantum numbers  a quite good correspondence to the WD distribution is recovered. For the first time, this effect has been discussed in \cite{GP56} in application to nuclear reactions for which quantum numbers might not be known in advance.  


A famous example is also given by the Bunimovich billiard \cite{B79} for which there are four independent energy subsets due to the symmetry of the boundary with respect to reflections in both vertical and horizontal directions. Correspondingly, there are four kinds of eigenstates specified by their symmetric properties in the configuration space. Thus, only by selecting a particular subset of energy levels related to a specific symmetry of the eigenstates, the WD distribution can be observed. Interesting enough, one can expect that for a $N-$dimensional Bunimovich billiard with $N \gg 1 $ the level spacing distribution $P(s)$ should be very close to the Poisson distribution when considering the total spectrum, and to the WD distribution when analyzing one of the subsets associated with a specified  symmetry of the eigenstates. As one can see, the question about the type of $P(s)$ characterizing the spectrum statistics of a given system is, strictly speaking, meaningless, unless all conditions are specified. It should be also noted that, even if the corresponding classical system is completely ergodic and chaotic (as in the case of the Bunimovich billiard) in the lower part of the energy spectrum the quantum effects always suppress chaos that prevents the  occurrence of WD distribution. The specific transition from a Poisson to a WD distribution as a function of the energy and the geometric shape of the billiard have been investigated in Ref.~\cite{BCL96} for the slightly perturbed Bunomivich billiard.

The emergence of the Poisson form of the  LSD widely treated as an indication of integrability, has attracted much attention from the viewpoint of its mechanism. Indeed, the integrability of a quantum system is closely related to regular sequences of energy levels. On the other hand, the Poisson distribution itself is known to appear in statistical physics as a strong property of randomness. The source of apparent randomness for the LSD has been studied, for the first time, in \cite{BT77}, where the emergence of the Poisson distribution was explained within a semi-classical approach to quantum systems with a well defined classical limit. Note that for one-dimensional systems the  LSD  is highly non-generic (and typically far from the Poisson). So far, the Berry-Tabor conjecture \cite{BT77} of the Poisson form of $P(s)$, as a generic property of quantum systems that are integrable in the classical limit, has not yet rigorously proved, in spite of the intensive mathematical studies done (see, for instance, \cite{S88,S91,M01} and references therein).


In more detail, the mechanism of  pseudo-randomness of the LSD  was firstly demonstrated in Ref.\cite{CCG85} in the numerical study of the rectangular billiard which is trivially integrable in both classical and quantum description. Obviously, despite the regularity of the energy spectrum, $E_{n,m} = \alpha n^{2} + m^{2}$ (with an irrational value of $\alpha$ and integers $n, m$) the LSD  was shown to have quite good correspondence with the Poisson distribution. Thus, the mechanism of apparent pseudo-randomness of the energy spacings has here a geometric nature, emerging due to the reduction of the two-dimensional set of the values $E_{n,m}$ to a one-dimensional set of $E$. Specifically, in spite of a regular grid of values $E_{n,m}$ on the plane ${n,m}$, the number of  points $E_{n,m}$ within the area bounded by the curves $E$ and $E+ds$ changes  randomly when changing slowly the value of $E$. On the other hand, strong deviations from the Poisson have been detected in the region of very small values of $s-$spacings \cite{CCG85}. Moreover, other sensitive statistical tests of the randomness of energy levels (such as the absence of correlations between distant energy levels) have shown that the sequence of energy levels cannot be considered truly random. These data indicate that statistical properties of energy spectra of integrable models have a restricted correspondence to the properties associated with a truly Poisson process. 


In this paper, we focus on the properties of energy spectra of the paradigmatic Lieb-Liniger (LL) model \cite{LL63,L63,Girardeu}, to which a huge number of works are devoted (see, for example, \cite{F17,Zo17,YYX15,Daviespra15} and references therein). This model describes one-dimensional (1D) bosons  on a circle interacting  with a two-body point-like interaction. The model belongs to a peculiar class of quantum integrable models solved by the Bethe ansatz \cite{Bethe, Korepin}; in particular,  it is possible to show that it has an infinite number of conserved quantities. Apart from the theoretical interest, many related problems have been recently discussed in view of various experiments with atomic gases~\cite{Go01,Po04,KWW04}. Since this model has no classical counterpart, it is extremely interesting to shed light on the mechanism for the emergence of randomness, if any, in the energy spectrum, and quantify its statistical properties. Moreover, using the Bethe equation it is possible to extract an arbitrary large number of energy levels within an arbitrary small numerical precision. This fact renders the LL model really unique in the class of interacting integrable  quantum many-body systems.

For a weak inter-particle interaction the LL model can be described in the mean-field (MF) approximation. Contrarily, for a  strong interaction, the 1D atomic gas enters the so-called Tonks-Girardeau (TG) regime in which the density of the interacting bosons becomes identical to that of non-interacting fermions (keeping, however, the bosonic symmetry for the wave function) \cite{Girardeu}. The crossover from one regime to the other is governed by the ratio  between the boson density $n$ and the interaction strength  $g$. The latter constant is inversely proportional to the 1D inter-atomic scattering length and can be experimentally tuned with the use of the Feshbach resonance (see, for example, \cite{Olshanii} and references therein).  

\section{The model} 
The Hamiltonian of the LL model with $N$ bosons interacting on a ring of length $L$ by a point-like interaction, has the form
\begin{equation}
\label{eq:LL}
H=H_0 + c V= -\sum_{i} \frac{\partial^2}{\partial x_i^2} +  2c \sum_{i\neq j} \delta(x_i-x_j).
\end{equation}
Here we have used the units in which $\hbar/2m = 1$ and the key parameter $c$ stands for the strength of the $\delta$-like interaction between bosons. 

For the reader convenience we report here the standard procedure for finding eigenvalues and eigenfunctions, see for instance Refs.\cite{F17,Zo17,YYX15,Daviespra15}.
The solution for the eigenvalue problem can be obtained firstly by restricting the configuration space to the sector $x_1 \leq x_2\leq \ldots \leq x_N$ where the Bose wave function $\Psi (x_1,\ldots,x_N)$ is completely determined. Thus, the system becomes a system of free particles with the interaction playing a role only as a boundary condition, for any $k=2,\ldots N$, 
$$
\left[ \frac{\partial \Psi}{\partial x_k} - \frac{\partial \Psi}{\partial x_{k-1}} \right]_{x_{k}=x_{k-1}} =
c\Psi.
$$

Now, we  search the $\Psi $-function (in the sector defined above) written in the form  
\begin{equation}
\label{eq:psi-a}
\Psi (x_1,..,x_N) = \sum_{P} a_P \exp\left[i\sum_{k=1}^{N} x_k \lambda_{P(k)}\right]
\end{equation}
where $a_P$ are  phase factors and the sum is over the $N!$ permutations of $1,..,N$. Imposing the latter to be a solution of the stationary Schrodinger equation allows one to get the phase factors in terms of the rapidities $\lambda_{P(k)}$.

Then, by fixing periodic boundary conditions on the circle of length $L$, namely, 
$\Psi(...,x_k,...)= \Psi(...,x_k+L,...)$, one obtains a system of $N$  Bethe equations (see, for example, \cite{F17}), $i=1,...,N$,
\begin{equation}
\label{eq:BE}
\lambda_i = \frac{2 \pi}{L} m_i - \frac{2}{L} \sum_{k \neq i}^{N} \arctan\left(\frac{\lambda_i-\lambda_k}{c}
\right)  
\end{equation}
for the rapidities $\lambda_i$.
Each set of distinct "quantum numbers" $\{m_i\}_{i=1}^N$ is composed by integers (or half integers)  for   an odd (or even) number of particles $N$.

Extending the wave function to the whole configuration space, each set of $N$ rapidities $\{\lambda_i\}_{i=1}^N$
defines an eigenstate $\ket{\alpha(\lambda_1,..., \lambda_N)}$. The set of all eigenstates serves as a complete basis in the completely symmetric many-body Hilbert space. We label with $\alpha$ the $N$ rapidities related to the eigenstates $\ket{\alpha}$ in the following way:  $\{\lambda_i^\alpha\}_{i=1}^N$.

The rapidities completely specify  the energy of the eigenstate $\ket{\alpha}$
\begin{equation}
\label{eq:E}
E(\lambda_1^\alpha,...,\lambda_N^\alpha) = \sum_{i=1}^N (\lambda_i^\alpha)^2,
\end{equation}
the momentum
\begin{equation}
\label{eq:P}
P(\lambda_1^\alpha,...,\lambda_N^\alpha) = \sum_{i=1}^N \lambda_i^\alpha
\end{equation}
and also the infinite set of conserved charges with $k>2$.
\begin{equation}
\label{eq:charge}
Q_k(\lambda_1^\alpha,...,\lambda_N^\alpha) = \sum_{i=1}^N (\lambda_i^\alpha)^k.
\end{equation}
In the following for simplicity we restrict our study by taking an odd value of $N$ and setting $L=2\pi$.

The goal of this paper is to find all possible eigenvalues in a finite energy region and to study in a very accurate way their statistics.   In order to do that we first fix a large integer number $M$ and consider all possible sets of $N$ different integers $|m_j|<M$. For each set we compute all possible sets of rapidities (that determine the energies) satisfying Eqs.~\ref{eq:BE}. Let us call $E_{max}(M)$ the maximal value of the energy for a given value $M$ of the cut-off. After, we consider another cut-off number $M^\prime > M$ and compute again all possible rapidities and energies. In this way we obtain more energy levels in the region with the maximal value $E_{max}(M') > E_{max}(M)$, but what is important is that we obtain many missing energy levels in the interval $[0,E_{max}(M)]$. We continue the procedure of increasing $M$ up to the complete filling of the interval $[0,E_{max}(M)]$, meaning that a further increase of $M$ does not produce any new eigenvalue in the specified energy range. Typically, we have found that $M^\prime \sim 2M$ is enough to find all eigenvalues in the interval  $[0,E_{max}(M)]$.

In order to solve the non-linear equations Eqs.~\ref{eq:BE}, we have used the standard Newton solver with the precision $\epsilon = 10^{-15}$ in finding the rapidities. As initial conditions we provided an educated guess taking into account that we know explicitly the solution in two simple limit cases, 
\begin{itemize}
	\item infinite interaction (fermionization, free fermions), 
        \begin{equation}
        \label{eq:cinf}
        c=\infty, \ \ \  \lambda_j^\alpha = \frac{2 \pi}{L} m_j^\alpha
        \end{equation}
	\item no interaction (free bosons),	
	\begin{equation}
        \label{eq:c0}
	 c=0, \ \ \  \lambda_j^\alpha = \frac{2 \pi}{L}\left(m_j^\alpha -j + \dfrac{N+1}{2}\right) .
	\end{equation}
\end{itemize}
Since any eigenenergy is the sum of $N$ squared distinct integer numbers (see Eq.(\ref{eq:E})) it is obvious that the distribution $P(s)$ is dramatically different from a Poisson in both limits above. 
\begin{figure*}[!ht]
    \centering 
    \includegraphics[width=\textwidth]{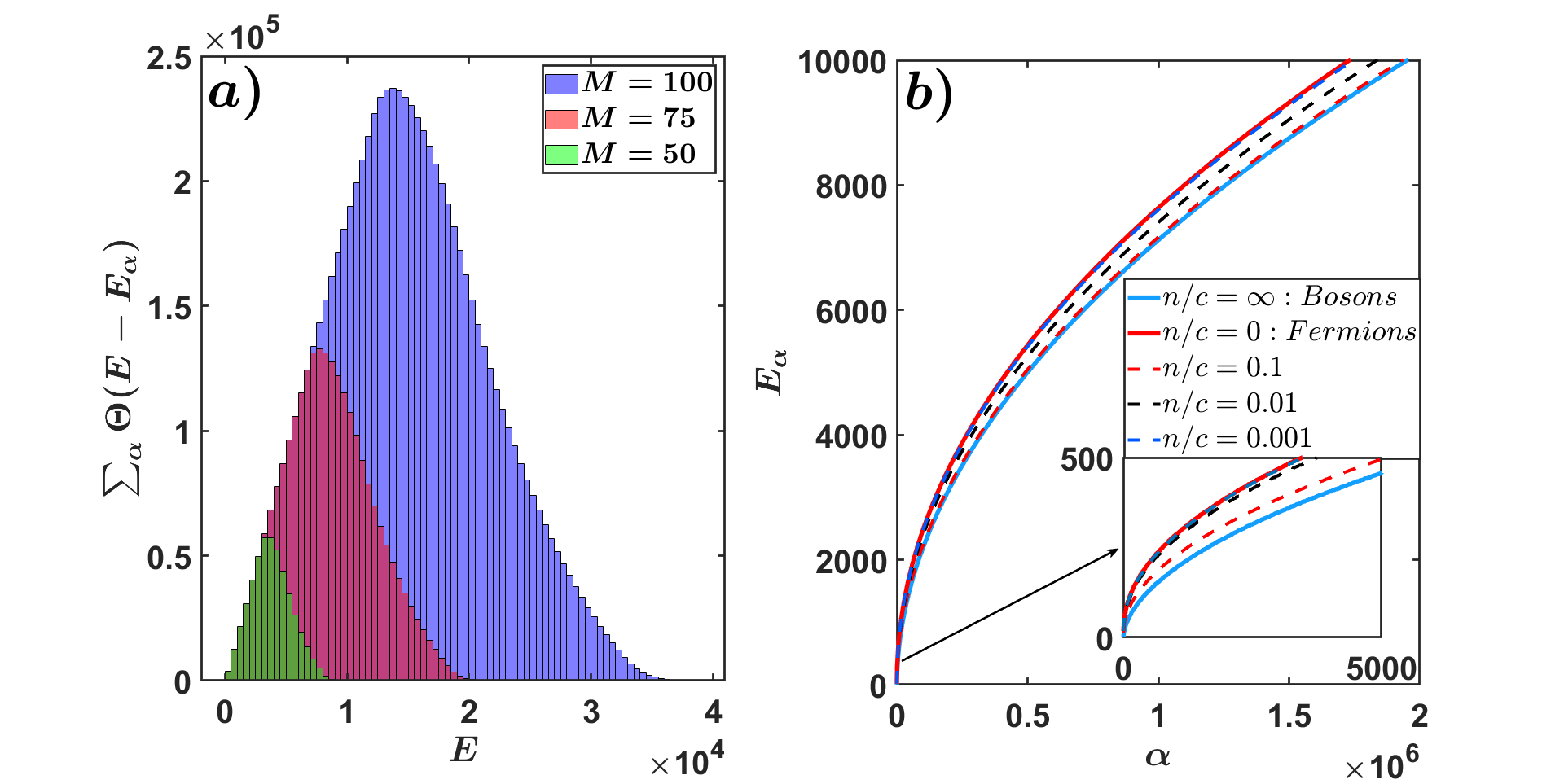}
    \caption{(a) Density of states for different values of the cut-off $M=50,75,100$ as indicated in the legend and medium interaction $n/c=0.01$. (b) Eigenvalues $E_\alpha$  as a function of the index label $\alpha$ for different interactions strength $n/c$ as indicated in the caption, together with the analytical solutions for infinite and zero interaction. 
Other parameters are $N=5$, $P=2$.
}
	\label{fig:f0}
\end{figure*}

As an example of spectrum, we compute all energy eigenvalues by taking three different values of $M$. Results for density of states are shown in Fig.~\ref{fig:f0}(a). As one can see the spectrum is linear in energy, with the bell shape which is simply due to the cut-off $M$. It is clear that in the infinite case ($M\to \infty$) only the linear unbounded spectrum remains. Needless to say,  we consider the eigenvalues taken from the linear part of the energy spectrum only, in which the values of energies are not suffered by the cut-off. An example of the spectrum, as a function of the label $\alpha$ and for different interaction strengths $n/c$ is given in 
Fig.~\ref{fig:f0}(b).
  
\section{Independent spectra}


\begin{figure*}[ht!]
	\centering
	\includegraphics[width=\textwidth]{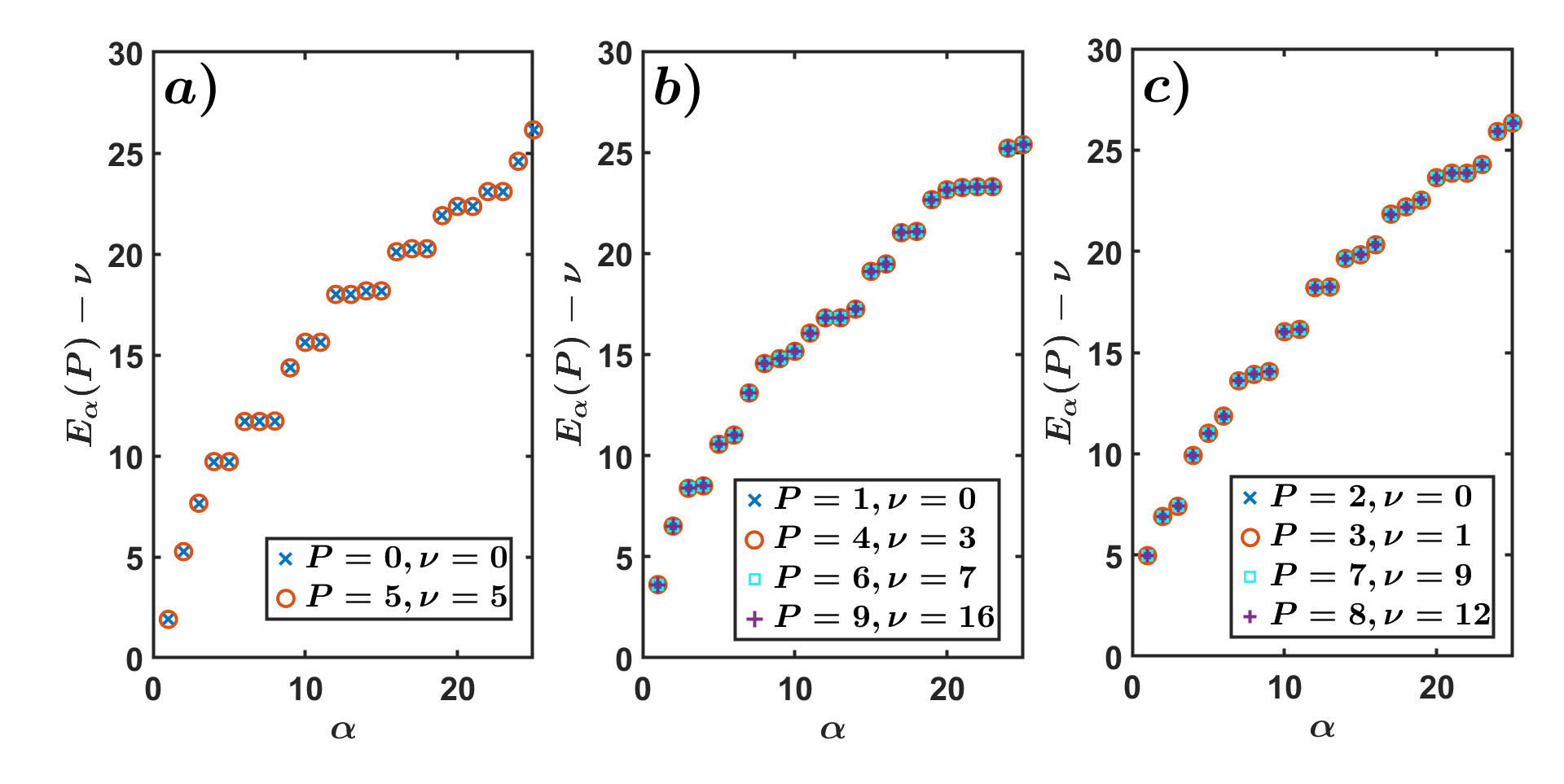}
	\caption{(a,b) First $25$ energies rescaled by the factor $\nu$ with interaction $n/c=1$ for $N=5$ particles for different starting momenta: a)$P=0$, b) $P=1$ and c) $P=2$.
}
	\label{fig:nu}
\end{figure*}


Before starting to analyze the spectral properties, let us consider the fundamental role played by the total momentum $P$. Since the $\arctan$ in Eq.~(\ref{eq:BE}) is an odd function of its argument, the total momentum for our choice ($L=2\pi$ and $N$ odd) is an integer number, 
$$
P =\sum_{i=1}^N \lambda_i = \sum_{i=1}^N m_i
$$
Arranging the eigenvalues according to  i) its momentum and for each fixed momentum according to  its growing energy, one can see that the Hamiltonian matrix has an infinite block diagonal structure. Each block matrix has been obtained by bracketing eigenstates having the same total momentum $P$, and it is disconnected from any other block (due to the momentum conservation there are no matrix elements connecting states with different total momenta). Moreover, a careful  analysis has led to the unexpected result indicating that the energy levels in different energy subsets corresponding to different values of the total momentum $P$, are strongly correlated. In order to show that let us first start from Eq.(\ref{eq:BE}) written as
\begin{equation}
\label{eq:BE1}
\lambda_i = m_i + \sum_{k \neq i} f(\lambda_i - \lambda_k) ,
\end{equation} 
with $ f(x)\equiv \arctan(x)/\pi $. Here, each set of different integers $\{ m_i \}^N_{i=1}$ determines a set of rapidities $\{ \lambda_i \}^N_{i=1}$ characterizing completely an eigenstate with the energy $E(\lambda_1,...,\lambda_N)$ and the momentum $P(\lambda_1,...,\lambda_N)$, see Eq.~(\ref{eq:E}). 

Let us now consider another (shifted) set of quantum numbers $m'_i = m_i + k$ with $k$ a positive or negative integer number. It is clear that the shifted rapidities $\lambda'_i = \lambda_i + k $ satisfy the same Eq.~(\ref{eq:BE1}),
\begin{equation}
\label{eq:BE2}
\lambda'_i = m'_i + \sum_{k \neq i} f(\lambda'_i - \lambda'_k) ,
\end{equation} 
but with a shifted  momentum and  energy given respectively by,
\begin{equation}
\label{eq:pp}
P \{ \lambda'_i \} = \sum_i (\lambda_i + k) = P + k N 
\end{equation}
and
\begin{equation}
\label{eq:pe}
\begin{array}{lll}
E \{ \lambda'_i \} &= \sum_i (\lambda_i + k)^2 =  E \{ \lambda_i \} + \nu ,
\end{array}
\end{equation}
where $\nu= 2 k P + k^2 N $ is an integer number.

This means that for a given number $N$ of particles, all the energies corresponding to some  fixed momentum $P$, turn out to be shifted by the same constant integer number $\nu$ (and thus the levels statistics for the energy subset with fixed total momentum, will be the same). In particular, let us note that for $k=-2P/N$ (consider here that only $k$ integer is valid) we have $\nu=0$ and $P \{ \lambda'_i \} = - P\{ \lambda_i \}$. Thus, we recover the fact that $P$ and $-P$ are related to the same eigenenergy. 

Let us now analyze Eqs.~(\ref{eq:pp},\ref{eq:pe}) in more detail. Setting, for instance, $k=1$ the energy spectrum for $P$ and $P+N$ is simply shifted by the factor $\nu=2P+N$. This suggests that, at most, only the spectra for $P=0,1,...N-1$ might be independent. However, it is not the case. Actually, the energy subsets for $P=1$ and $P=N-1$ have the same spectrum (with a constant shift). To see this, one has to simply take $k=1$ and $P=-1$ in Eq.~(\ref{eq:pe}) and to observe that $P=1$ and $P=-1$ give the same spectrum. In the same way the energy subsets for $P=2$ and $P=N-2$ are the same with respect to a constant shift, and so on. The bottom line is that for an odd number of particles $N$ only the spectra with the momentum values  $P=0,1,...,(N-1)/2$ are independent, all the other being simply shifted by a constant. This is a quite unexpected property of the energy spectra since it is completely independent of the interaction strength $c$.

A numerical verification of the above mathematical proof is shown in Fig. \ref{fig:nu}.  There, we present the first $100$ energy levels for the LL model with $N=5$ bosons for the chosen rescaled (shifted) value $n/c=1$ of the interaction. The energy levels have been obtained by solving the Bethe  equations for few values of the total momentum, namely, for $0 \le P \le 9$, and plotted together with the energies corresponding to the momenta $P=0,1,2$. The data demonstrate that all eigenvalues, properly shifted by factor $\nu$, are exactly the same.


\begin{figure*}[ht!]
	\centering
	\includegraphics[width=\textwidth]{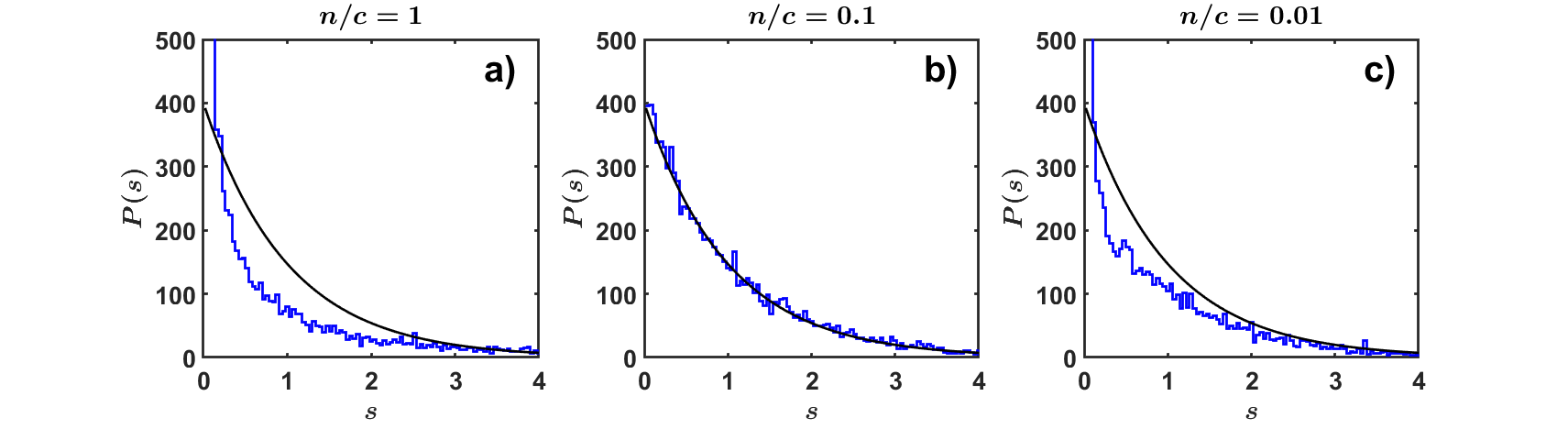}
	\caption{LSD distribution for $N=5$ particles, fixed momentum $P=2$ and $n/c=1$ (a), $n/c=0.1$ (b) and $n/c=0.01$ (c). Statistics have been obtained taking the first $10^4$ different energy levels starting from the beginning of the energy spectrum at fixed momentum. Black curves indicate the Poisson distribution.
}
	\label{fig:nnls}
\end{figure*} 

\section{Statistics of close energies}

As we explained above, one can exactly compute an amount of eigenvalues in a given energy region by choosing a large enough value of the cutoff $M$. In this way we will be able  to explore the statistical properties of energy spectra without the influence of spectrum truncation, at least for a  not very large number of particles and for not very high energies. 

One of the conclusions from our study is that it is meaningless to expect the Poisson form of the LSD when speaking of the total energy spectrum. Namely, from our analysis it is clear that there is a strong clustering of energy levels at $s=0$ that does not allow to speak properly about  the Poisson distribution. There are two mechanisms of such clustering. The first is the influence of a high degeneracy of energy levels in two limit cases (either zero or infinite inter-particle interaction. And the second mechanism of clustering is due to very strong correlations between the subsets of energy levels, belonging to different values of the total momentum (see the proof above). 

Much more interesting is the question about the statistical properties of energy spectra for a fixed total momentum. In this situation there is only the first mechanism of clustering for a relatively weak or very strong interaction. Let us consider for instance the LSD for the first $10^4$ energy levels for three different interaction strengths. Results are shown in Fig.~\ref{fig:nnls}. As one can see, the clustering of energy levels for small values of $s$ persists both for a weak interaction, $n/c = 1$, and for a very strong interaction, $n/c = 0.01$ as one can clearly see in panels (a) and (c) of Fig.~\ref{fig:nnls}. Here, we use the rescaled parameter $n/c$ of the interaction, which was found to be the key characteristic for the crossover from to the mean-field to Tonks-Girardeau regime.  In both cases one can see that  the LSD has a very pronounced peak at the origin, at variance with the Poisson distribution shown for comparison. Also, a clear similarity between these two cases is clearly seen. Due to the highly non-generic feature of the LSD with strong clustering at $s=0$ occurring for the total energy spectrum, in what following we address a much more interesting question about the spectrum statistic for fixed values of the total momentum. As for the intermediate interaction strength, one can see that, formally, the LSD that can be treated as the Poisson one. We did not explore this distribution more carefully (for example with the use of the $\chi^2$ test), instead, we have used other famous tests for the check of the absence of correlations between nearest energy levels.  

Thus, concerning the Poisson form of the LSD in the Lieb-Liniger model, one can conclude that it can be observed under some conditions only (for fixed total momentum, plus not weak or strong inter-particle interaction). It should be stressed that our results are restricted by a relatively small number of particles. Moreover, it is an open question what happens with an increase of the energy, specifically, it is not clear whether the form of the LSD changes or not. In any case, 
a widespread opinion that the level spacing distribution in completely integrable quantum models is always of the Poisson form, is not true (at least, for the Lieb-Liniger model).  

Let us now discuss another test recently suggested for the discriminating the Poisson statistics from the Wigner-Dyson one. Specifically, we focus on the ratio of consecutive level spacings that is used in the literature (see, for instance, \cite{OH07,ABGR13,CDK14,CR20} and references therein) when analyzing the repulsion between nearest levels. This test has the advantage of not requiring the unfolding of the spectrum since it involves the two closest energy levels only. Following this approach, we introduce the variables
\begin{equation}
\label{eq:chi}
\xi_n=\dfrac{s_n}{s_{n-1}} \ \ \ , \ \ \ s_n=E_n-E_{n-1} 
\end{equation} 
and create the quantity of our interest,
\begin{equation}
\label{eq:chimin}
\chi_n= \ \ {\rm Min}(\xi_n, \ 1/\xi_n)
\end{equation}
Even if the spectrum unfolding is not required, this test is really powerful  when an average over disorder is used, focusing on a particular part of the energy spectrum. According to the theory, supported by various numerical studies, in case of the Poisson distribution for energy spacing one gets, $\langle \chi_n \rangle = 0.386$.

\begin{figure*}[ht!]
	\centering
	\includegraphics[width=\textwidth]{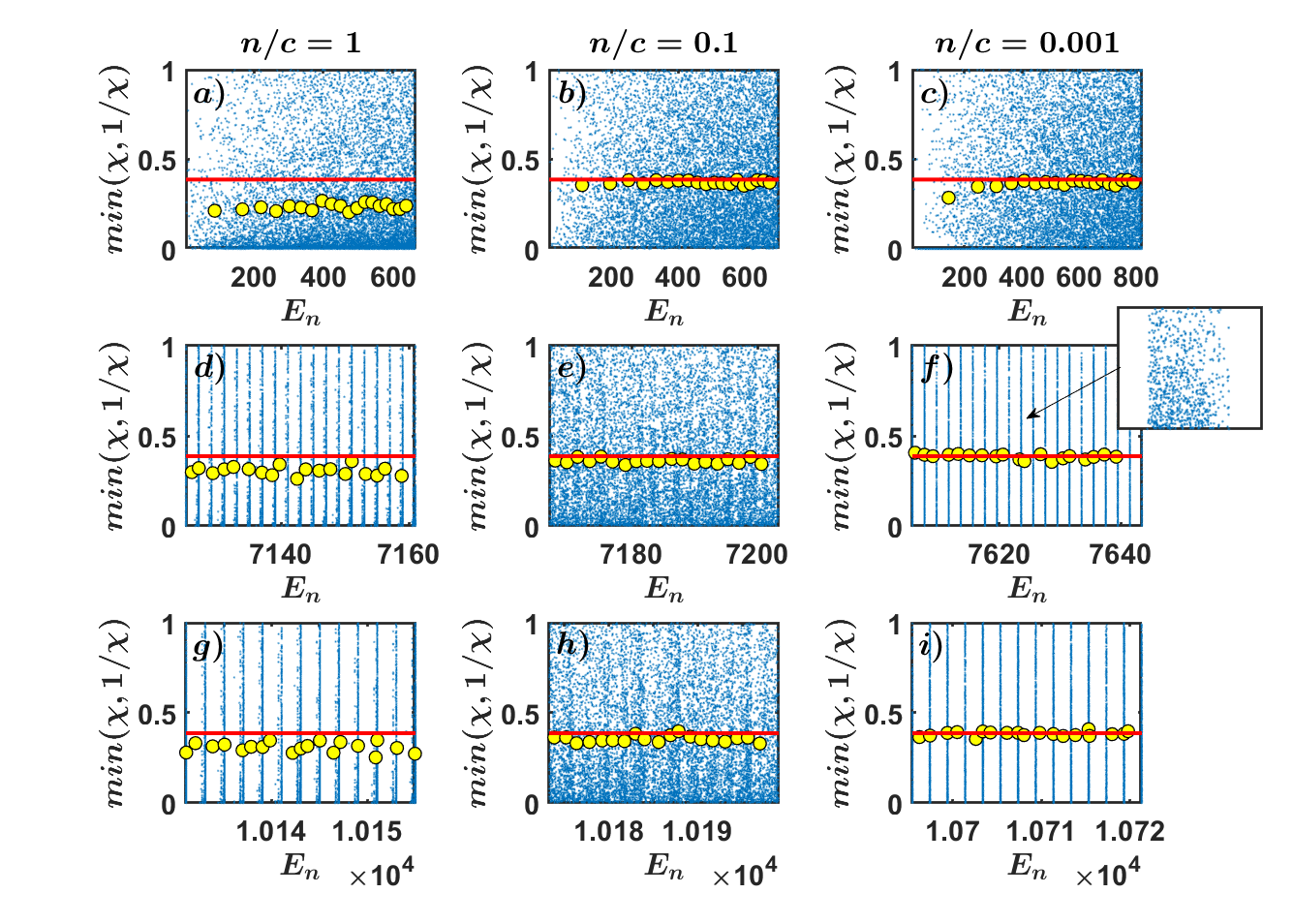}
	\caption{Blue dots: $10^4$ values of $\min(\chi_n \ , \ 1/\chi_n)$ for $N=5$ particles and $P=2$ of the total momentum. Panels in different rows stand for different parts of the energy spectrum (see x-axis), while different columns indicate  different values of the interaction strength (as indicated in the legend). Yellow dots are shown for the average over $500$ close values, while the continuous red line corresponds to the value $0.386$ obtained from a completely random set of energies. In panel (f) a blow up of a single blue vertical line is shown. }
	\label{fig:chi}
\end{figure*}

In Fig.\ref{fig:chi} we summarize our study of the above quantity $\langle \chi_n \rangle$ for different interaction strengths $n/c$, various values of the total momentum $P$ in different parts of the energy spectrum.  Our results (light blue dots) are compared to the theoretical value related to the Poisson distribution (red line). As one can see, the variable $\chi_n$ undergoes huge fluctuations which can hardly give a precise answer about the kind of the LSD. 
To facilitate the comparison we average $\chi_n$ over $500$ consecutive energy values (yellow circles).  The averaged data show that, in general,  there is a good agreement with the value corresponding to the Poisson distribution, obtained for the intermediate interaction between bosons. Moreover, as one can see from the middle-bottom right panels (high energy and strong interaction), even if the yellow circles are well fitted by the red line (indicating Poisson statistics) an extreme clustering of the levels indicates a complete absence of randomness as one should expect.

\section{$\Delta_3$ statistics}
Randomness in the energy spectrum can be detected by analyzing not only short-range correlations, e.g. nearest-neighbor statistics, but also long-range ones. According to this test, known as the $\Delta_3$-statistics, one can reveal the so-called {\it rigidity} (or stiffness) of the energy spectrum and discriminate between regular and chaotic motion in the corresponding classical systems \cite{SVZ84}. The strongest rigidity can be associated with that given by the equidistant energy levels. As was shown by Dyson and Mehta \cite{Dyson,M67}, the spectrum of full random matrices reveals a kind of rigidity which is due to correlations between distant energy levels. Such a rigidity of energy spectrum, can be compared with a slightly melted crystal, the analogy which has been used by Dyson to derive many statistical properties of random matrices.

 To imply this test for physical systems, it is necessary to proceed with the unfolding of the sequence of energy levels due to the dependence of the mean level spacing on the energy. The idea of the unfolding is to pass from a given sequence of levels to that having the constant level spacings, however, with the same correlations between the levels, both short and long-range ones.  
In this approach we start with the function,
\begin{figure*}[ht!]
	\centering
	\includegraphics[width=\textwidth]{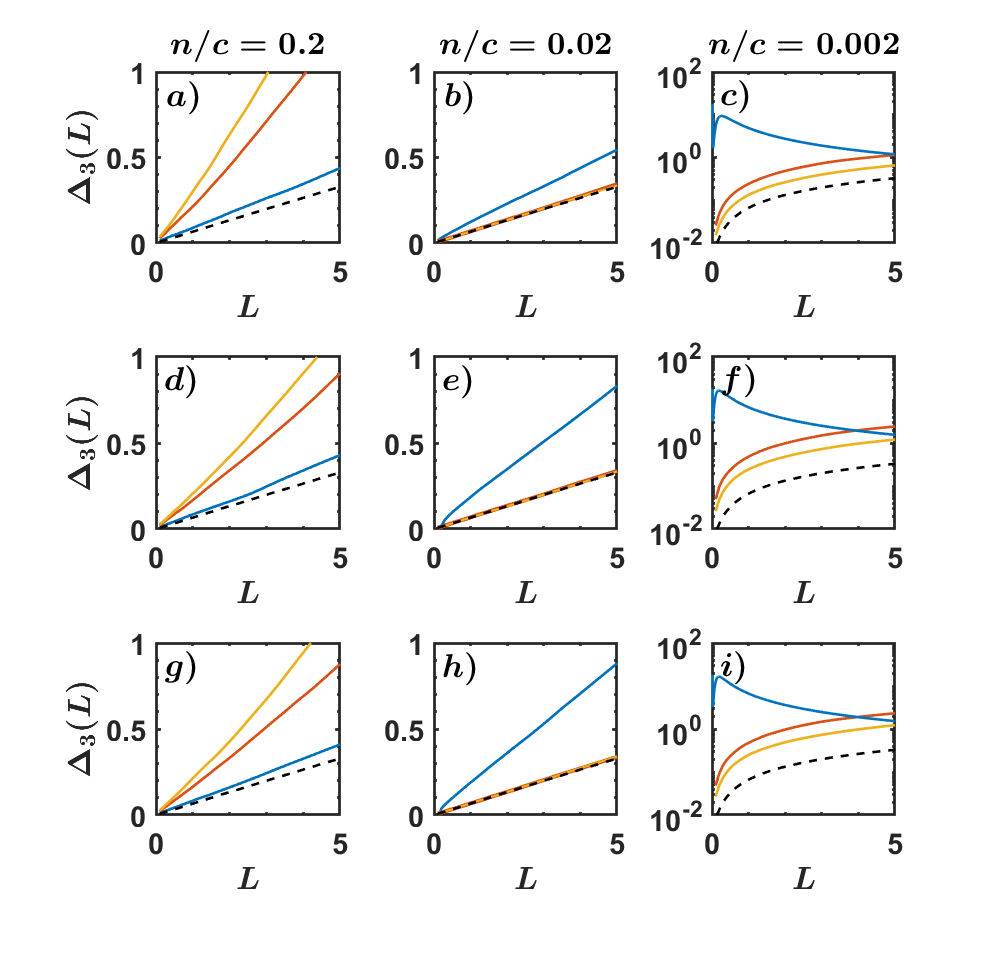}
        \caption{The $\Delta_3$ statistics obtained for different values of the total momentum, $P=0,1,2$ (respectively upper, middle and lower row), averaged over $10^4$ consecutive initial energies. Each column represents different interaction strengths:  $n/c=0.2$ (left),  $n/c=0.02$ (middle), $n/c=0.002$ (right). Different colors used for different energy regions: low energy region with $\xi_n$ starting from $n=1$ in blue, middle energy region with $\xi_n$ starting from 
        $n=10^6$ in red, and $\xi_n$ starting from $n=2 \times 10^6$ in yellow,
        Dashed black line stands for the Poisson statistics. In computing spectra for the case $P=0$ the accidental degeneracy of few eigenvalues has been eliminated.
}
	\label{fig:d3}
\end{figure*}

\begin{equation}
\label{eq:eta}
\eta(E)=\sum_{n=1}^{N} \Theta(E-E_n)
\end{equation}
which counts the number of levels with the energy less than or equal to $E$ and is usually referred to as the staircase function. Specifically, the unfolding consists in  mapping the sequence $\{E_1,E_2,...,E_N\}$ onto the numbers $\{\xi_1,\xi_2,...,\xi_N\}$ in such a way that the function $\xi(E)$ is the smooth part of $\eta(E)$ and $\hat{\eta}_{fl}(E)$ is the fluctuating part: $\eta(E)=\xi(E)+\hat{\eta}_{fl}(E)$. Thus, we define,
\begin{equation}
\label{eq:d3}
\Delta_3 = min_{A,B} \frac{1}{L} \int_{\xi_s}^{\xi_s+L} [\hat{\eta}(\xi) - A \xi - B]^2 d\xi 
\end{equation}
where $\hat{\eta}(E)$ counts the number of levels in the interval $[\xi_s,\xi_s+L]$. Minimizing (\ref{eq:d3}) we obtain the following relations,
\begin{equation}
\begin{cases} \frac{d \Delta_3}{dA} = - \frac{2}{L} \int_{\xi_s}^{\xi_s+L} \xi [\hat{\eta}(\xi) - A \xi - B] d\xi = 0 \\ \frac{d \Delta_3}{dB} = - \frac{2}{L} \int_{\xi_s}^{\xi_s+L} [\hat{\eta}(\xi) - A \xi - B] d\xi = 0 \end{cases},  
\end{equation}
from which one finds,
\begin{equation}
\label{eq:ab}
\begin{cases} A = \dfrac{p x_1 - 2q}{x_1^2 - 2 x_2} \\ B = \dfrac{q x_1 - p x_2}{x_1^2 - 2 x_2} \end{cases} 
\end{equation}
where
\begin{equation}
x_1 = \frac{2}{L} \int_{\xi_s}^{\xi_s+L} \xi  d\xi = 2 \xi_s + L ,
\end{equation}
\begin{equation}
x_2 = \frac{2}{L} \int_{\xi_s}^{\xi_s+L} \xi^2 d\xi = \frac{2}{3} (L^2 3 \xi_s^2 + 3 \xi_s L) ,
\end{equation}
\begin{equation}
p = \frac{2}{L} \int_{\xi_s}^{\xi_s+L} \hat{\eta}(\xi) d\xi \\, \, \,\,\,\, q = \frac{2}{L} \int_{\xi_s}^{\xi_s+L} \xi \hat{\eta}(\xi) d\xi ,
\end{equation}
and
\begin{equation}
t = \frac{2}{L} \int_{\xi_s}^{\xi_s+L} \hat{\eta}^2(\xi) d\xi .
\end{equation}

With these parameters the expression for $\Delta_3$ takes the form,
\begin{equation}
\label{eq:d3sol}
\Delta_3 = \frac{1}{2} t + \frac{1}{2} A^2 x_2 + B^2 - A q - B p + A B x_1
\end{equation} 
In case of a Poisson statistics the relation $\Delta_3(L)=  L/15$ holds. In Fig.~\ref{fig:d3} we show the $\Delta_3$ statistics for $N=5$ and all possible values of total momentum $P=0,1,2$ giving rise to independent spectra (one line for each fixed momentum). Moreover, we considered three different values of the inter-particle interaction, $n/c=0.2; 0.02; 0.002$ (from the left to the right column). In each panel we show the dependence of $\Delta_3$ on $L$ for not too large values $L<5$, obtained in three different energy regions of the spectrum: a low energy region close to the ground state (in blue), a middle (red) and a high one (orange). Below we extend our analysis to the larger $L$ values. As one can see from the comparison with the analytical prediction (indicated by  a dashed line),  the $\Delta_3$ test shows some similarity with the Poisson statistics apparently  only for intermediate values of the interaction (middle column) and sufficiently high energy values (not too close to the ground state).

\begin{figure}[t]
	\centering
	\includegraphics[width=8cm]{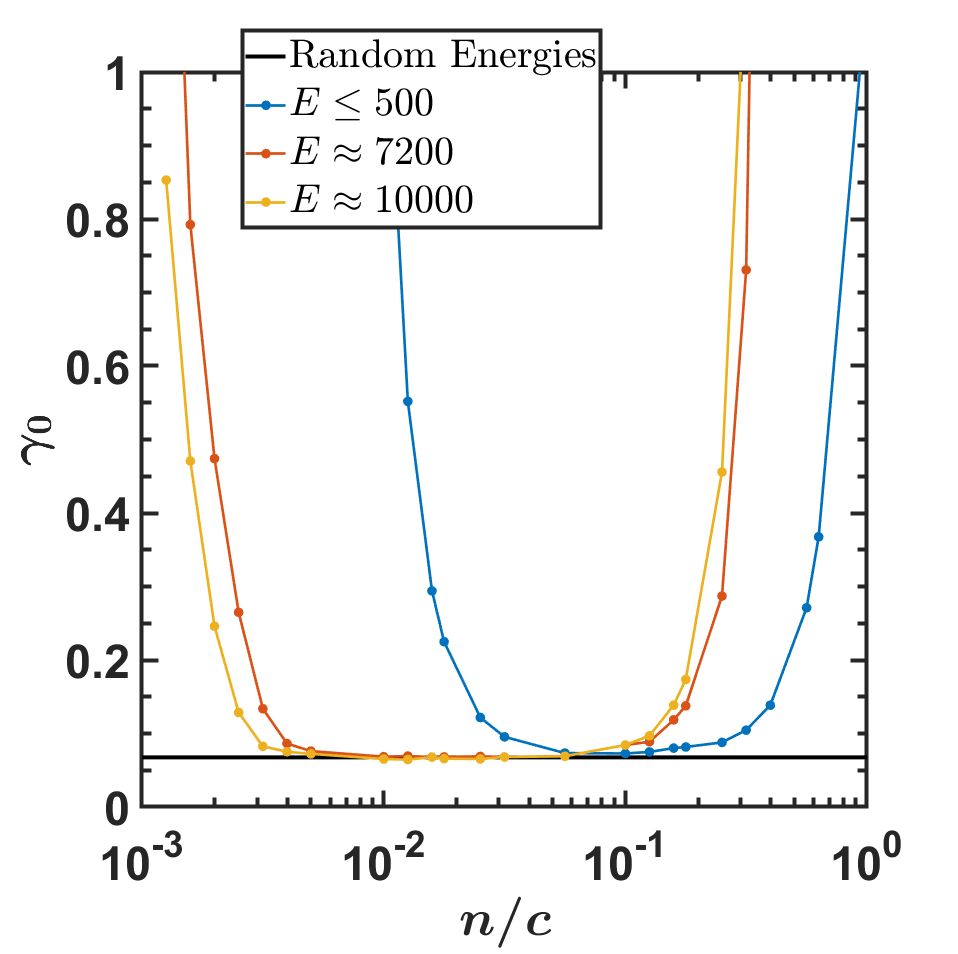}
	\caption{Slope of $\Delta_3(L)$ for different values of the interaction and location in the spectrum as in previous figures, compared with the slope obtained with a completely random sequence (see the black line). The data are presented for $P=2$ total momentum, and the average is done over a set of $10^4$ energies around the energy indicated in the legend. The $\log$-scale in $x$-axis shows a nice symmetry with respect to a particular interaction strengths $n/c$ in dependence on the chosen energy range.   
}
	\label{fig:slope}
\end{figure}

In order to check quantitatively the deviations from the prediction of the Poisson process (straight line with the slope $1/15$ ) we fit $\Delta_3(L)$  to a line,
\begin{equation}
\Delta_3 (L) = \gamma_0 L + \gamma_1 ,
\end{equation} 
in the range $0 \le L \le 5$. Then we plot in in Fig.~\ref{fig:slope} the slope $\gamma_0$ as a function of the interaction strength $n/c$ for different energy regions in the spectrum (see different colors). From this picture it is clear that from one side one can say that for any energy range a suitable range of values of the interaction can be found where the $\Delta_3$ test indicates a rather good agreement with the Poisson statistics (in the region $0<L<5$). On the other hand, for any interaction strength one can find an energy range where the $\Delta_3$ statistics indicates strong deviations from the Poisson predictions.
\begin{figure*}[t]
	\centering
	\includegraphics[width=\textwidth]{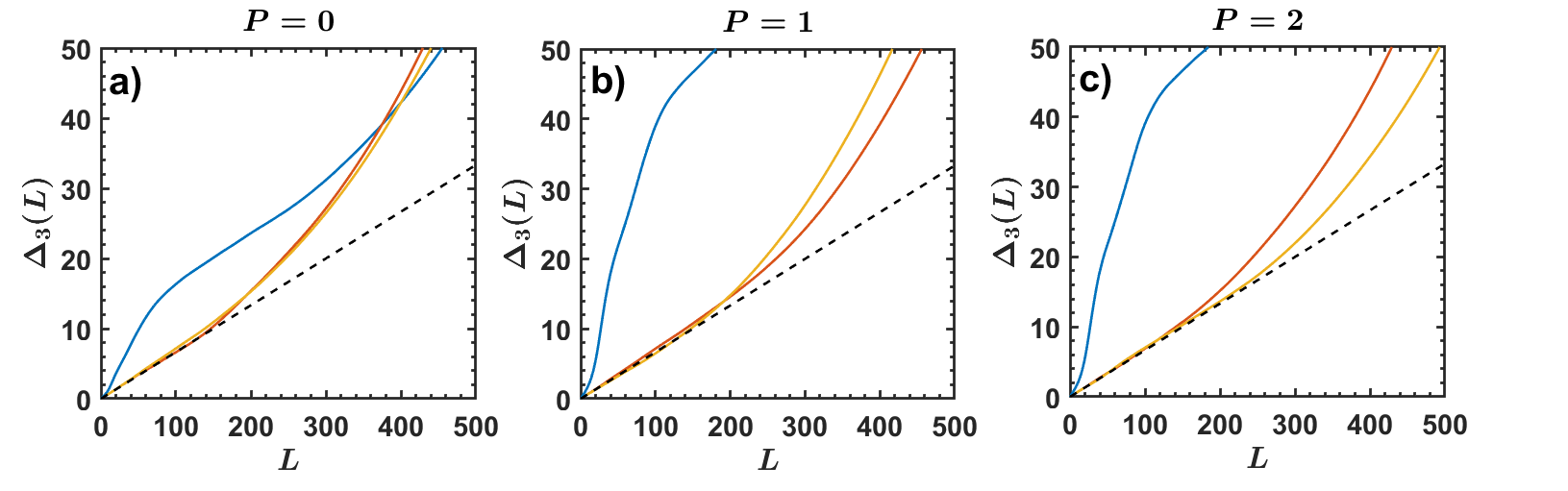}
        \caption{Average $\Delta_3$ statistics for large $L$ values. The parameters are the same as in Fig.\ref{fig:d3}, central column
}
	\label{fig:largeL}
\end{figure*} 
Finally, we extend our study of the $\Delta_3$ statistics on a scale much larger than $L=5$. As was found in Ref.~\cite{CCG85} for the (integrable) rectangular billiard, with an increase of $L$ the deviation from the Poisson statistics is increasing, even if both the LSD and $\Delta_3$ tests have shown quite good correspondence to the theoretical predictions. Specifically, it was found that a kind of stiffness of the energy levels reveals when  very large values of $L$ are taking into account. Thus, here we ask the question whether the correlations between the energy levels not seen in our study for the intermediate values of interaction, will emerge on a larger $L-$ scale. To this end we study the $\Delta_3$ statistics for  very large $L$-values for those interaction strength values for which the Poisson statistics seems to confirm the theoretical predictions (see, for instance, the middle column of Fig.~\ref{fig:d3}). Results for this study are shown in Fig.~\ref{fig:largeL}. As one can see, the linear rigidity of the Poisson spectrum lasts approximately up to $L \lesssim 200$ (we took into account a statistics using $10^4$ different energy levels). Beyond this scale, we have discovered a dramatic increase for the slope of the $\Delta_3$ statistics. This effect is similar to that found numerically in Ref.\cite{CCG85}. However, to our surprise the deviations for the slope occur in the opposite direction. Namely, at variance with the results of Ref.\cite{CCG85} demonstrating a kind of a decrease of the rigidity, our data clearly indicate an {\it increase of the rigidity} of the energy spectrum with an increase of $L$. Specifically, the slope of $\Delta_3$ statistics increases with $L$ (we remind that for the Wigner-Dyson statistics, $\Delta_3(L) \approx \ln(L)$) and not decreases as found in Ref.\cite{CCG85}. This fact that $\Delta_3 \propto L^\alpha$ with $\alpha >1$ for $L\gg 1$ which we have observed numerically, may have a strong impact for the dynamical properties of the Lieb-Liniger model, and still awaits for additional clarifications.

\section{Conclusions and discussion}
\begin{itemize}
	\item In this paper we have studied, both analytically and numerically, the statistical properties of energy spectra of the famous Lieb-Liniger (LL) model. This model describes a number of bosons moving on a ring and interacting via a point-like interaction. It is known to be completely integrable due to the Bethe ansatz approach and, thus, allows for the analytical studies. In recent years, the problem of the statistical description of many-body integrable models has attracted much attention, both from the theoretical and experimental view points. Our interest to this model was restricted only to a particular question of the statistical properties of energy spectra, paying  main attention to the role of the conservation of the total momentum.  
	
	The main point of our analysis is that the properties of energy spectra of the LL-model dramatically depends on whether we consider the total energy spectrum or only a subset of energy levels associated with a fixed value of the total momentum (which is a physical constant of motion besides the energy). We have rigorously shown that the energy levels characterized by  different values of the total momentum are strongly correlated, thus leading to a  strong level clustering, independently on the strength of the inter-particle interaction. Specifically, the energy levels of some of the subsets can be obtained from other subsets of levels by a simple shift along the energy spectrum. As a result, concerning the total energy spectrum it is not possible to avoid strong clustering of energy levels. Therefore, the Poisson form of the level spacing distribution (LSD) does not appear in this situation. This effect of clustering which does not depend on the interaction, may have a strong impact for the dynamical properties of the model, and needs further analysis.  
 
	\item Having strong conclusion about the clustering of energy levels for the total energy spectrum, we considered the  properties of spectra for a fixed value of the total momentum. First, we have to note that for the zero and infinite inter-particle interaction there is a  strong degeneracy of the energy levels, that disappears moving away from both limits (therefore, for either weak or strong but finite interaction). In this case the LSD may approach the Poisson distribution, depending on the interaction strength and the energy region. Our data, indeed, confirm that depending on the model parameters one can speak of the Poisson form of LSD, however, the residual effect of the clustering of levels for $s=0$ may remain. Still, the question about what happens with an increase of the number of particles or with an increase of the energy, awaits for a further clarification.
	
	We have also used another test to explore the absence of correlations between nearest energy levels, namely the test based on the ratio of consecutive level spacings. In contrast with the LSD test, it does not need the unfolding of the energy spectrum. However,  the considered ratios of spacings reveal very strong fluctuations that can be only washed out by a further averaging over a large numbers of  spacings. Nevertheless, we have numerically overcame this problem and found a quite good confirmation of the results obtained by studying the LSD. Mainly, the predictions based on the  assumption of the Poisson process applied to the energy levels, are confirmed for the region of relatively strong interaction and for high enough energies.
	
	\item Our further analysis is due to the so-called $\Delta_3$ statistic, the test which measures the correlations between distant energy levels, that give an information about the rigidity of the spectrum. This test is much more sensitive when compared with the results  predicted by the {\it Poisson process} which provides a completely random sequence of levels. We have to remind that the well known test of the Poisson form of the LSD is a particular property of the Poisson process, exploring the correlations between the nearest levels in the energy sequence and leaving aside the question about long-term correlations. One of the first applications of the $\Delta_3$ statistics to the energy spectrum, was the study of statistical properties of energy spectra of the trivially integrable 2D rectangular billiard \cite{CCG85}. First, it was shown that when the global shape of the LSD looks like a Poisson, the careful study of the statistics of small level spacings ($s \ll 1$) can manifest serious deviations from the true Poisson dependence. Second, in the situation when one can accept a quite good correspondence of the LCD to the Poisson, the $\Delta_3$ test shows a strong deviation from the predicted linear dependence, $\Delta_3 \sim L \gg 1$ for $L \gg 1$. 
	
	To this end, we have carefully studied the $\Delta_3$ statistics in the LL-model and found that this test survives small values of $L$ (in the situations when the LSD meets the test of the Poisson dependence). However, with an increase of $L$ one can observe strong deviations from the theoretical prediction of the Poisson process. It is interesting that these deviations contrast  to those found in Ref. \cite{CCG85}. Specifically, at variance with the results for the rectangular billiard for which $\Delta_3 \sim L^{\alpha}$ with $\alpha < 1$, for the LL-model we have found $\alpha > 1$ (for $L \gg 1$ ). 
	
    \item Finally, we would like to stress that our numerical study is free from the effects of the cutoff of the energy spectrum. 
    As is known, any numerical approach to physical systems with an infinite energy spectrum suffers from a kind of cutoff of the spectrum. Thus, the control of the accuracy of numerically obtained energy levels is a typical problem in any numerical approach (see, for instance, \cite{RDYO07,JKB09,Daviespra15,Zo17,MBI20}). The peculiarity of our approach is such that the accuracy of the computation of energy levels is determined by the numerical solver of the equations Eq. \ref{eq:BE} determining the rapidities $\lambda$ only. Since the accuracy in solving these equations is extremely high, one can treat our computation as the {\it exact} one. As for the cutoff $M$ when specifying the integer numbers $m_j$ (see Section II), we found the way to avoid the influence of this cutoff by choosing the energy regions for which an increase of $M$ does not lead to any correction (see details in Section II). 
\end{itemize} 

{\em Acknowledgements.}--
We acknowledge discussions with G.~L. Celardo. F.B. acknowledges support by the Iniziativa Specifica I.N.F.N.-DynSysMath. FMI acknowledges financial support from CONACyT (Grant No. 286633).


\begin{thebibliography} {99}

\bibitem{berry} M.~V. Berry, Quantizing a classically ergodic system: Sinai's billiard and the KKR method, Ann. Phys. (N.Y.) {\bf 131}, 163 (1981).

\bibitem{valz} G. Casati, F. Valz-Gris, I. Guarneri, On the connection between quantization of nonintegrable systems and statistical theory of spectra, Lett. Nuovo Cimento {\bf 28}, 279 (1980).

\bibitem{BGS} O. Bohigas, M.~J. Giannoni, C. Schmit,  Characterization of chaotic quantum spectra and universality of level fluctuation laws, Phys. Rev. Lett. {\bf 52}, 1 (1984).

\bibitem{CCIF79} G.~Casati, B.~V. Chirikov, F.~M.~Izraelev, and J.~Ford, {\em Stochastic Behavior of Classical and Quantum Hamiltonian Systems}, Lect. Notes in Phys. {\bf 93}, Edited by G.~Casati and J.~Ford (Springer, Berlin, Heidelberg, 1979), p. 334.

\bibitem{bill} S.~W. McDonald and A.~N. Kaufman, Spectrum and eigenfunctions for a Hamiltonian with stochastic trajectories, Phys. Rev. Lett. {\bf 42}, 1189 (1979).

\bibitem{R92} L.E. Reichl, The Transition to Chaos, Springer-Verlag, New-York, 1992.

\bibitem{WVFS90} H. Wu, M. Vallires, Da Hsuan Feng, and D.W.L. Sprung,
Guassian-orthogonal-ensemble level statistics in a one-dimensional system, Phys. Rev. A {\bf 42}, 1027 (1990).

\bibitem{BJS03} L. Benet, F. Leyvraz, and T.H. Seligman, Wigner-Dyson statistics for a class of integrable models, Phys. Rev. E {\bf 68}, 045201(R) (2003).

\bibitem{BJL03} L. Benet, C. Jung and F. Leyvraz, Integrability of interacting two-level boson systems, J. Phys. A: Math. Gen.
{\bf 36}, L217 (2003).

\bibitem{FM09} F.W.K. Firk  and S. J. Miller, Nuclei, Primes and the Random Matrix Connection, Entropy {\bf 1}, 64 (2009).

\bibitem{G39} I. I. Gurevich, Nature, {\bf 144}, 326 (1939).

\bibitem{W51} E. Wigner, On the statistical distribution of the widths and spacings of nuclear resonance levels, Proc. Cambridge Phil. Soc. {\bf 47} 790 (1951).

\bibitem{HH58} J. Harvey and D. Hughes, Spacings of nuclear energy levels, Phys. Rev. {\bf 109}, 471 (1958).

\bibitem{Wigner} E.P. Wigner, Characteristic vectors of bordered matrices with infinite dimensions, Ann. of Math. {\bf 62}, 548 (1955); Characteristics vectors of bordered matrices with infinite dimensions II, Ann. of Math. {\bf 65}, 203 (1957); On the distribution of the roots of certain symmetric matrices, Ann. of Math. {\bf 67}, 325 (1958).

\bibitem{Dyson} F.J. Dyson, Statistical theory of the energy levels of complex systems I, J. Math. Phys. {\bf 3}, 140 (1962); Statistical theory of the energy levels of complex systems II, J. Math. Phys. {\bf 3}, 157 (1962); Statistical theory of the energy levels of complex systems III, J. Math. Phys. {\bf 3}, 166 (1962).

\bibitem{GMW98} T. Guhr, A. Müller–Groeling, and H. A. Weidenmüller, Random-matrix theories in quantum physics: common concepts, Phys. Rep., {\bf 299} 189 (1998).

\bibitem{GP56} I. I. Gurevich, and M. I. Pevzner, Repulsion of nuclear levels. Physica {\bf 22}, 1132 (1956).

\bibitem{B79} L. A. Bunimovich, On the Ergodic Properties of Nowhere Dispersing Billiards, Commun. Math Phys., {\bf 65} 295 (1979).

\bibitem{BCL96} F.Borgonovi, G.Casati and B.Li, Diffusion and Localization in Chaotic Billiards,
Phys. Rev. Lett. {\bf 77} , 4744, (1996)

\bibitem{PR60} C. Porter and N. Rosenzweig, Repulsion of energy levels in complex atomic spectra, Phys. Rev. {\bf 120} 1698 (1960).

\bibitem{BT77} M. V. Berry and M. Tabor, Level Clustering in the Regular Spectrum, Proc. R. Soc. Lond. A 1977 {\bf 356}, 375 (1977).

\bibitem{S88} Ya. G. Sinai, The absence of the Poisson distribution for spacings between quasi-energies in the quantum kicked-rotator model, Physica D, {\bf 33} 314 (1988).

\bibitem{S91} Ya. G. Sinai, Mathematical problems in the theory of quantum chaos, In: Lindenstrauss J., Milman V.D. (eds) Geometric Aspects of Functional Analysis. Lecture Notes in Mathematics, vol 1469. Springer, Berlin, Heidelberg.

\bibitem{M01} J. Marklof, Proceedings of the XIIIth International Congress on Mathematical Physics, London 2000 (International Press, Boston, 2001), pp. 359-363.

\bibitem{CCG85} G. Casati, B. V. Chirikov and I. Guarneri, Energy-level statistics of integrable quantum systems, Phys. Rev. Lett., {\bf 54} 1350 (1985). 


\bibitem{LL63} E. H. Lieb and W. Liniger, Exact Analysis of an Interacting Bose Gas. I. The General Solution and the Ground State, Phys. Rev. {\bf 130}, 1605 (1963).

\bibitem{L63} E. H. Lieb, Exact Analysis of an Interacting Bose Gas. II. The Excitation Spectrum, {\bf 130}, 1616 (1963).

\bibitem{Girardeu}  M. Girardeau, Relationship between Systems of Impenetrable Bosons and Fermions in One Dimension, J. Math. Phys. {\bf 1}, 516 (1960).

\bibitem{F17} F. Franchini, An Introduction to Integrable Techniques for One-Dimensional Quantum Systems, Lecture Notes in Physics 940, (2017).

\bibitem{Zo17} J. C. Zill, T. M. Wright, K. V. Kheruntsyan1, T. Gasenzer, M. J. Davis, Quantum quench dynamics of the attractive one-dimensional Bose gas via the coordinate Bethe ansatz, SciPost Phys. {\bf 4}, 011 (2018). 

\bibitem{YYX15} Jiang Yu-Zhu, Chen Yang-Yang,  and  Guan Xi-Wen, Understanding many-body physics in one dimension from the Lieb–Liniger model, Chin. Phys. B {\bf 24} 050311 (2015).

\bibitem{Daviespra15} J. C. Zill, T. M. Wright,  K. V. Kheruntsyan,  T. Gasenzer,  and M.J. Davis, Relaxation dynamics of the Lieb-Liniger gas following an interaction quench:
A coordinate Bethe-ansatz analysis, Phys. Rev. A, {\bf 91}, 023611 (2015)

\bibitem{Bethe}  H. A. Bethe, Zur Theorie der Metalle, Z. Phys {\bf 71}, 205 (1931).

\bibitem{Korepin}  V. E. Korepin, N. M. Bogoliubov, and A. G. Izergin, Quantum Inverse Scattering Method and Correlation Functions (Cambridge University Press, Cambridge, UK, 1993).

\bibitem{Go01} A. Gorlitz et al., Realization of Bose-Einstein Condensates in Lower Dimensions, Phys. Rev. Lett. {\bf 87}, 130402 (2001). 

\bibitem{Po04} B. Paredes et al., Tonks–Girardeau gas of ultracold atoms in an optical lattice, Nature (London) {\bf 429}, 377 (2004).

\bibitem{KWW04} T. Kinoshita, T. Wenger, and D.S. Weiss, Observation of a one-dimensional Tonks-Girardeau gas, Science {\bf 305}, 1125 (2004), 

\bibitem{Olshanii} M. Olshanii, Atomic Scattering in the Presence of an External Confinement and a Gas of Impenetrable Bosons, Phys. Rev. Lett. {\bf 81}, 938 (1998).

\bibitem{OH07} V. Oganesyan and D.A. Huse, Localization of interacting fermions at high temperature,  Phys. Rev. B {\bf 75}, 155111 (2007). 

\bibitem{ABGR13} Y. Y. Atas, E. Bogomolny, O. Giraud, and G. Roux, Distribution of the Ratio of Consecutive Level Spacings in Random Matrix Ensembles, Phys. Rev. Lett. {\bf 110}, 084101 (2013).

\bibitem{CDK14} N.D. Chavda, H.N. Deota, and V.K.B. Kota, Poisson to GOE transition in the distribution of the ratio of consecutive level spacings, Physics Letters A {\bf 378} (2014) 3012.

\bibitem{CR20} A. L. Corps and A. Relano, Distribution of the ratio of consecutive level spacings for different symmetries and degrees of chaos, Phys. Rev. E {\bf 101}, 022222 (2020).

\bibitem{SVZ84} T. H. Seligman, J.J.M. Verbaarschot, and M.R. Zirnbauer, Quantum Spectra and Transition from Regular to Chaotic Classical Motion, Phys. Rev. Lett. {\bf 53}, 215 (1984). 

\bibitem{M67} M. L. Mehta, Random Matrices (Academic, New York, 1967).

\bibitem{MBI20} S. Mailoud, F. Borgonovi, and F.M. Izrailev, Process of equilibration in many-body isolated systems: diagonal versus thermodynamic entropy, New J. Phys. {\bf 22} 083087 (2020).

\bibitem{RDYO07} M. Rigol, V. Dunjko, V. Yurovsky, and M. Olshanii, Relaxation in a Completely Integrable Many-Body Quantum System: An Ab Initio Study of the Dynamics of the Highly Excited States of 1D Lattice Hard-Core Bosons, PRL {\bf 98} 050405 (2007).

\bibitem{JKB09} D. Jukić, B. Klajn, and H. Buljan, Momentum distribution of a freely expanding Lieb-Liniger gas, Phys. Rev. A {\bf 79} 033612 (2009).


 




\end{thebibliography}
\end{document}